# Measuring a Kaluza-Klein radius smaller than the Planck length

Frank Reifler and Randall Morris

Lockheed Martin Corporation, MS2 137–205
199 Borton Landing Road
Moorestown, NJ 08057

**Abstract:** Hestenes has shown that a bispinor field on a Minkowski space-time is equivalent to an orthonormal tetrad of one-forms together with a complex scalar field. More recently, the Dirac and Einstein equations were unified in a tetrad formulation of a Kaluza-Klein model which gives precisely the usual Dirac-Einstein Lagrangian. In this model, Dirac's bispinor equation is obtained in the limit for which the radius of higher compact dimensions of the Kaluza-Klein manifold becomes vanishingly small compared with the Planck length. For a small but finite radius, the Kaluza-Klein model predicts velocity splitting of single fermion wave packets. That is, the model predicts a single fermion wave packet will split into two wave packets with slightly different group velocities. Observation of such wave packet splits would determine the size of the Kaluza-Klein radius. If wave packet splits were not observed in experiments with currently achievable accuracies, the Kaluza-Klein radius would be bounded by at most $10^{-25}$ times the Planck length.

# 1. Introduction

Using geometric algebra, Hestenes showed in 1967 that a bispinor field on a Minkowski space-time is equivalent to an orthonormal tetrad of one-forms together with a complex scalar field, and that fermion plane waves can be represented as isometric modes of the tetrad [1]. More recently, the Dirac and Einstein equations were unified in a tetrad formulation of a Kaluza-Klein model which gives precisely the usual Dirac-Einstein Lagrangian [2], [3]. In this model, the self-adjoint modes of the tetrad describe gravity, whereas, as in Hestenes' work, the isometric modes of the tetrad together with a scalar field describe fermions. An analogy can be made between the tetrad modes and the elastic and rigid modes of a deformable body [2]. For a deformable body, the elastic modes are self-adjoint and the rigid modes are isometric with respect to the Euclidean metric on $R^3$. This analogy extends into the quantum realm since rigid modes satisfying Euler's equation can be Fermi quantized [4]. As with Euler's equation for a rigid body, the tetrad formulation of Dirac's partial differential bispinor equation is a classical Hamiltonian system, with (noncanonical) unitary Lie-Poisson brackets [4]. Fermi quantization of such classical systems is possible whenever the Lie algebra can be represented by fermion creation and annihilation operators. Note that most Lie algebras can be represented by fermion operators [5], so there exist many classical Lie-Poisson systems which can be Fermi quantized.

The use of tetrads to describe gravity has a long history [6], which includes coupling with the Dirac field as a source [7]. However, introducing a tetrad to describe both fermion and gravitational fields solves an important problem posed by current theories of fermion-graviton interaction. To define bispinors, reference tetrad fields or their equivalent must be defined on the space-time manifold [8]. However, only ten of the sixteen components of a tetrad field describe gravity. The remaining six components are supernumerary boson fields in current gravitational theories [9]. In the Kaluza-Klein tetrad model, the tetrads, which do not require a reference field, describe both fermions and gravity without superfluous degrees of freedom [2].

The tetrad Kaluza-Klein model is based on a constrained Yang-Mills formulation of the Dirac Theory [2], [3], [4], [10], [11], [12]. In this formulation a bispinor field $\Psi$ is mapped to a set of SL(2,R) × U(1) gauge potentials $A_\alpha^K$ and a complex scalar field $\rho$. The map $\Psi \to (A_\alpha^K, \rho)$ imposes an orthogonal constraint on the gauge potentials $A_\alpha^K$. Apart from the exceptional set $\rho = 0$, the map $\Psi \to (A_\alpha^K, \rho)$ is a double covering map onto its image. (Such a double covering map has no observable effects [4], [10], [13].) The image of this map contains precisely the gauge potentials $A_\alpha^K$ which satisfy the orthogonal constraint:

$$A_\alpha^K A_{K\beta} = -|\rho|^2 g_{\alpha\beta} \tag{1.1}$$

where $g_{\alpha\beta}$ denotes the space-time metric. The gauge index K = 0,1,2,3 is lowered and raised using a gauge metric $g_{JK}$ and its inverse $g^{JK}$ (see Section 2). Repeated indices are summed. We show in formula (2.11) in Section 2 that via the map $\Psi \to (A_\alpha^K, \rho)$, the Dirac bispinor Lagrangian (2.1) equals a constrained Yang-Mills Lagrangian in the limit of an infinitely large coupling constant which we denote as $g_0$.

In the Kaluza-Klein formulation of the tensor Dirac theory, we map the fermion field $(A_\alpha^K, \rho)$ to a tetrad of vector fields $v_K$ and a complex scalar field, also denoted as $\rho$, on a smooth manifold M = X × G, where X is a space-time and G = SL(2,R) × U(1) (see Section 3). The tetrad $v_K$ together with a (fixed) basis of right-invariant vector fields on G determines a metric denoted as $\langle\,,\,\rangle$, a volume form denoted as $d\gamma$, and also a curvature two-form denoted as R( , ), on M (see Section 3). The unified action S for the gravitational and fermion fields is given by

$$S = \int L\, d\gamma \tag{1.2}$$

where the unified Lagrangian, L, is (see Section 3):

$$L = \frac{1}{16\pi\kappa_0} R_v + \frac{1}{g_0} \overline{v_K(\rho+\mu)} v^K(\rho+\mu) \qquad (1.3)$$

where $\kappa_0$ is Newton's gravitational constant, $g_0$ is the Yang-Mills coupling constant referred to previously, and $\mu = \frac{2m_0}{g_0}$, where $m_0$ is the fermion mass. In formula (1.3), we employ the sum of sectional curvatures restricted to the subspace spanned by the tetrad $v_K$:

$$R_v = \sum_{J=0}^{3} \sum_{K=0}^{3} \langle R(v_J, v_K) v^J, v^K \rangle \qquad (1.4)$$

By formulating the Kaluza-Klein Lagrangian (1.3) with the tetrad $v_K$, the orthogonal constraint (1.1) is eliminated (see Section 3).

The limit on the Yang-Mills coupling constant $g_0$ has a geometric significance in the Kaluza-Klein tetrad model, in that as $g_0$ becomes infinitely large, as required to obtain the usual Dirac-Einstein equations from the Lagrangian (1.3), the radius of the higher compact dimensions in the Kaluza-Klein model becomes vanishingly small, even when compared to the Planck length [14]. This can be seen from the following argument. In the Lagrangian (1.3) the constants $g_0$, $\kappa_0$, and $\mu$ are functions of three "fundamental" constants, $m_0$, $\delta_0$, and $\lambda_P$, where $m_0$ is the fermion mass, $\delta_0$ is a radius which characterizes the size of the higher compact dimensions of the Kaluza-Klein manifold M, and $\lambda_P$ is the Planck length. In Section 3 we show that

$$\delta_0 = \left(\frac{8\pi}{g_0^3}\right)^{1/2} \lambda_P \qquad (1.5)$$

Thus, in the limit required to obtain Dirac's equation, as $g_0$ becomes infinitely large, $\delta_0$ is much smaller than the Planck length $\lambda_P$.

For nonvanishing values of the radius $\delta_0$, the Dirac equation obtained from the Lagrangian (1.3) is nonlinear (in the bispinor variables $\Psi$), and solutions of this equation exhibit a phenomenon known as velocity splitting, whereby a free fermion wave packet splits into two wave packets traveling with a small velocity difference [15], [16]. In Section 4 we shall derive formulas relating $g_0$ to the velocity splitting in free fermion wave packets.

Thus in principle it is possible from formula (1.5) to determine or bound the radius $\delta_0$ with fermion beam experiments designed to detect velocity splitting in wave packets. Consider a current experiment where single electrons are emitted at 100 kilometer intervals in wave packets of length $10^{-5}$ meters, traveling over a meter at half the speed of light [17]. From formulas (4.16) and (4.17) in Section 4, assuming that velocity splitting is not observed, we can estimate that $g_0$ must be greater than $10^{17}$, and thus from formula (1.5), $\delta_0$ must be smaller than $10^{-25}$ times the Planck length. Experiments with slower electrons or with protons could reduce the above bound on $\delta_0$ by twenty orders of magnitude.

These experiments can be performed at a "first quantized" level with single fermions and in the absence of a discernible gravitational field. The reason is that from formula (1.5), a non-vanishing radius $\delta_0$ determines a small fermion self-interaction constant $1/g_0$ in terms of which the generalized Dirac Hamiltonian H can be written as:

$$H = H_0 + \frac{1}{g_0} H_1 \qquad (1.6)$$

where $1/g_0 \approx 10^{-17}$ is a very small dimensionless parameter, $H_0$ is exactly the usual Dirac bispinor Hamiltonian, and both $H_0$ and $H_1$ are integrals of measurable functions of the bispinor field $\Psi$. By the Spectral Theorem both $H_0$ and $H_1$ (after regularization common in quantum field theories) can be represented as self-adjoint operators, and thus, a perturbative quantum field theory could be formulated as for other nonlinear fields, with $1/g_0$ as the expansion parameter.

Although the practical use of such a nonlinear theory is very difficult, the Lagrangian (1.3) has observable predictions at the classical or first quantized level for the nonlinear wave phenomena discussed in Section 4. Such predictions do not conflict with quantum field theory because only the fermion part of the Lagrangian (1.3) contains the radius $\delta_0$, and this radius is only manifested as the small dimensionless coupling constant $1/g_0$. (It is shown in formula (3.33) of Section 3 that the Lagrangian (1.3) is the sum of the usual Hilbert-Einstein Lagrangian for the gravitational field plus a Yang-Mills Lagrangian for the fermion field.) Since $\delta_0$ only slightly perturbs the fermion wave packets, quantum effects at the Planck length scale, such as the effect of gravity fluctuations on the nonlinear fermion wave packets, would not be observable in the experiments proposed in this paper [8]. (Even the earth's gravity as an external field would not be discernible in the proposed experiments.)

Therefore, at the first quantized level there is no conflict with quantum field theory in proposing an experiment with freely propagating, single fermions in order to observe a small self-interaction of the Dirac equation. Also, the classical or first quantized equations in this paper are then sufficient to derive observable predictions from the Kaluza-Klein model.

We conclude this introduction with some brief remarks on the implications of the tetrad Kaluza-Klein model. While it is generally agreed that the classical limit for (a large number) of photons is the classical electromagnetic field, it is also widely believed that no classical limit exists in the same sense for fermions [9], [18], [19]. We believe that this belief is unfounded given that, as previously discussed, fermions, gravitons, and gauge bosons can be unified at a classical level in a tetrad Kaluza-Klein model [3]. Also the observability of the higher dimensions of the tetrad Kaluza-Klein model through velocity splitting, suggests new experiments to test quantum mechanics in a nonlinear regime.

In Section 2 of this paper we review the derivation which demonstrates that the Dirac bispinor Lagrangian equals a constrained Yang-Mills Lagrangian in the limit of an infinitely large coupling constant. We show how all bispinor observables are directly derived from well known Yang-Mills formulas. Then in Section 3 we show how both the limit and the orthogonal constraint (1.1) are explained geometrically in a Kaluza-Klein tetrad model. Finally, in Section 4 we show how the Kaluza-Klein radius $\delta_0$ can be measured in velocity splitting experiments.

## 2. Tensor form of the Dirac Lagrangian

In previous papers we derived the tensor form of Dirac's bispinor Lagrangian and reviewed the history of such derivations by Takahashi and others [2], [4], [11]. To introduce the notation needed for the remainder of this paper, we will briefly review in this section the derivation which demonstrates that the Dirac bispinor Lagrangian (2.1) equals the constrained Yang-Mills Lagrangian (2.11) in the limit of an infinitely large coupling constant. (In Kaluza-Klein geometry this limit is equivalent to the radius of the higher compact dimensions being very small compared to the Planck length.) In addition, we will show how all bispinor observables (e.g., the energy-momentum tensor $T^{\alpha\beta}$, spin polarization tensor $S^{\alpha\beta\gamma}$, and the electric current vector $J^{\alpha}$ for the Dirac bispinor field) can be derived directly from well known Yang-Mills formulas.

Dirac's bispinor Lagrangian $L_D$ for the bispinor field $\Psi$, is given by

$$L_D = \text{Re}[i\overline{\Psi}\gamma^{\alpha}\partial_{\alpha}\Psi - m_0 s] \qquad (2.1)$$

where s is the complex scalar field defined by

$$\begin{aligned} \text{Re}[s] &= \overline{\Psi}\Psi \\ \text{Im}[s] &= i\overline{\Psi}\gamma^5\Psi \end{aligned} \qquad (2.2)$$

and where $\gamma^\alpha$ for $\alpha = 0, 1, 2, 3$ and $\gamma^5$ are Dirac matrices [20], $m_0$ denotes the fermion mass, $\partial_\alpha$ denote partial derivatives with respect to space-time coordinates, and (using bispinor notation) $\overline{\Psi} = \Psi^+ \gamma^0$, where $\Psi^+$ denotes the transpose conjugate of $\Psi$. Tensor indices $\alpha, \beta, \gamma$ are lowered and raised using the Minkowski space-time metric, which we denote as $g_{\alpha\beta}$, and its inverse $g^{\alpha\beta}$. Repeated tensor indices are summed from 0 to 3.

It was previously shown that except for the mass term, Dirac's bispinor Lagrangian (2.1) is invariant under $SL(2,R) \times U(1)$ gauge transformations [11]. Moreover, it was shown that the scalar s in formula (2.2) is invariant under SL(2,R) gauge transformations, and transforms as a complex U(1) scalar under the U(1) gauge transformations (i.e., chiral gauge transformations [11]). To make the Lagrangian (2.1) invariant for all $SL(2,R) \times U(1)$ gauge transformations, it was shown to suffice that $m_0$ transform like $\bar{s}$ (the complex conjugate of s). Since $m_0$ appears in the Lagrangian (2.1) without derivatives, the assumption that $m_0$ transform like $\bar{s}$ under U(1) chiral gauge transformations, has no effect on the Dirac equation [11].

Also as previously shown [11], from the Dirac bispinor Lagrangian (2.1) we can derive the following $SL(2,R) \times U(1)$ Noether currents $j_\alpha^K$ for K = 0,1,2,3. In particular, $j_\alpha^0$ is the electromagnetic current and $j_\alpha^3$ is the chiral current; i.e.,

$$\begin{aligned} j_\alpha^0 &= \overline{\Psi}\gamma_\alpha\Psi \\ j_\alpha^3 &= \overline{\Psi}\gamma_\alpha\gamma^5\Psi \end{aligned} \qquad (2.3)$$

whereas [11],

$$\begin{aligned} j_\alpha^1 &= \text{Re}[\overline{\Psi}\gamma_\alpha\Psi^C] \\ j_\alpha^2 &= \text{Im}[\overline{\Psi}\gamma_\alpha\Psi^C] \end{aligned} \qquad (2.4)$$

where $\Psi^C$ denotes the charge conjugate of $\Psi$. Note that $j_\alpha^0$, $j_\alpha^1$, and $j_\alpha^2$ are the SL(2,R) Noether currents, and $j_\alpha^3$ is the U(1) Noether current [11]. The Noether currents $j_\alpha^K$ and scalar s satisfy an orthogonal constraint known as a Fierz identity [11], [21], [22]:

$$j_\alpha^K j_{K\beta} = |s|^2 g_{\alpha\beta} \qquad (2.5)$$

where gauge indices J, K, L are raised and lowered using a Minkowski metric $g_{JK}$ (with diagonal elements $\{1,-1,-1,-1\}$ and zeros off the diagonal) and its inverse $g^{JK}$. As with space-time tensor indices, repeated gauge indices are summed from 0 to 3. Note from formulas (2.3) and (2.4) that the Noether currents $j_\alpha^K$ are real.

As shown previously [11], we can map the Noether currents $j_\alpha^K$ into a subset of SL(2,C) × U(1) currents $J_\alpha^K$ by setting:

$$J_\alpha^K = (J_\alpha^0, \mathbf{J}_\alpha) = (-j_\alpha^3, -ij_\alpha^2, ij_\alpha^1, -j_\alpha^0) \qquad (2.6)$$

where $\mathbf{J}_\alpha = \left(J_\alpha^1, J_\alpha^2, J_\alpha^3\right)$ are complex SL(2,C) currents and $J_\alpha^0$ is the U(1) current. We then map a subset of SL(2,C) × U(1) gauge potentials $A_\alpha^K$ and a complex scalar field $\rho$ into $(J_\alpha^K, s)$ by setting:

$$J_\alpha^K = 4|\rho|^2 A_\alpha^K$$

$$s = 4|\rho|^2 \bar\rho \qquad (2.7)$$

By formula (2.6) the gauge potentials $A_\alpha^K$ are restricted to an SL(2,R) × U(1) subgroup for which

$$\mathrm{Re}[A_\alpha^1] = \mathrm{Re}[A_\alpha^2] = \mathrm{Im}[A_\alpha^3] = 0 \qquad (2.8)$$

Note that from formulas (2.6) and (2.7) that $A_\alpha^0$ is real.

Using different notation Takahashi [22] derived the following formula for Dirac's bispinor Lagrangian (2.1):

$$L_D = -\mathrm{Re}[(\partial_\alpha \mathbf{A}_\beta) \bullet \mathbf{A}^\alpha \times \mathbf{A}^\beta + 2i\bar\rho A_\alpha^0 \partial^\alpha \rho + 4m_0 |\rho|^2 \bar\rho] \qquad (2.9)$$

where $\mathbf{A}_\alpha = \left(A_\alpha^1, A_\alpha^2, A_\alpha^3\right)$, with the orthogonal constraint (2.5) expressed as:

$$A_\alpha^K A_{K\beta} = -|\rho|^2 g_{\alpha\beta} \qquad (2.10)$$

(Formulas (2.9) and (2.10) are derived from first principles in reference [11].) Once the SL(2,C) × U(1) gauge symmetry of formula (2.9) is recognized, the demonstration that Dirac's bispinor Lagrangian (2.1) equals a constrained Yang-Mills Lagrangian in the limit of an infinitely large coupling constant, is fairly obvious. Consider the following Yang-Mills Lagrangian $L_g$ for the gauge potentials $A_\alpha^K$ and the complex scalar field $\rho$:

$$L_g = -\frac{1}{4g}\operatorname{Re}\left[A_{\alpha\beta}^K A_K^{\alpha\beta}\right] + \frac{1}{g_0}\overline{D_\alpha(\rho+\mu)}\,D^\alpha(\rho+\mu) \qquad (2.11)$$

where the Yang-Mills field tensor $A_{\alpha\beta}^K = \left(A_{\alpha\beta}^0, \mathbf{A}_{\alpha\beta}\right)$ is defined as:

$$A_{\alpha\beta}^0 = \partial_\alpha A_\beta^0 - \partial_\beta A_\alpha^0$$

$$\mathbf{A}_{\alpha\beta} = \partial_\alpha \mathbf{A}_\beta - \partial_\beta \mathbf{A}_\alpha - g\mathbf{A}_\alpha \times \mathbf{A}_\beta \qquad (2.12)$$

whereby the Yang-Mills coupling constant g is the self-coupling of the gauge potentials $\mathbf{A}_\alpha$. Furthermore, in the Lagrangian (2.11), the complex scalar $\mu$ satisfies:

$$\mu = \frac{2m_0}{g_0}, \qquad \partial_\alpha \mu = 0 \qquad (2.13)$$

where $m_0$ is the fermion mass, and $g_0 = (3/2)g$. As previously stated for Dirac's bispinor Lagrangian (2.1) both the complex scalar field s and the fermion mass $m_0$ transform as U(1) scalars. The same is true for $\rho$ and $\mu$ by formulas (2.7) and (2.13). Hence the covariant derivative $D_\alpha$ acts on $\rho+\mu$ as follows:

$$D_\alpha(\rho+\mu) = \partial_\alpha \rho + ig_0 A_\alpha^0(\rho+\mu) \qquad (2.14)$$

That is, $g_0 = (3/2)g$ is the Yang-Mills constant which couples the U(1) scalars $\rho$ and $\mu$ to the U(1) gauge potential $A_\alpha^0$. Then as previously shown [12], from formulas (2.9) through (2.14), Dirac's bispinor Lagrangian (2.1) equals:

$$L_D = \lim_{g\to\infty} L_g \qquad (2.15)$$

Note that the Euler-Lagrange equation for the Lagrangian (2.11) with the orthogonal constraint (2.10) expressed using Lagrange multipliers, commutes with the restriction (2.8). Hence, the $\mathbf{A}_\alpha$ can be used to denote either SL(2,C) or the subset of SL(2,R) gauge potentials. By regarding SL(2,R) as embedded in the complex analytic group SL(2,C), we are able to use familiar vector operations to express the Lie algebra structure constants in formulas (2.9) and (2.12). The vector operations greatly simplify derivations.

Note also from the Lagrangian (2.15) that we can derive all bispinor observables (e.g., the energy-momentum tensor $T^{\alpha\beta}$, spin polarization tensor $S^{\alpha\beta\gamma}$, and the electric current vector $J^{\alpha}$) directly from the Yang-Mills formulas. For example, the Dirac spin polarization tensor $S^{\alpha\beta\gamma}$ is usually expressed in bispinor notation as:

$$S^{\alpha\beta\gamma} = -\frac{1}{4}\overline{\Psi}(\gamma^{\alpha}\sigma^{\beta\gamma} + \sigma^{\beta\gamma}\gamma^{\alpha})\Psi \qquad (2.16)$$

where $\sigma^{\alpha\beta} = (i/2)(\gamma^{\alpha}\gamma^{\beta} - \gamma^{\beta}\gamma^{\alpha})$. Using the identity [7]:

$$\gamma^{\alpha}\sigma^{\beta\gamma} + \sigma^{\beta\gamma}\gamma^{\alpha} = 2\varepsilon^{\alpha\beta\gamma\delta}\gamma_{\delta}\gamma^{5} \qquad (2.17)$$

together with formulas (2.3), (2.6), (2.7), and (2.10), formula (2.16) reduces to:

$$S^{\alpha\beta\gamma} = -\frac{1}{2}\varepsilon^{\alpha\beta\gamma\delta}\overline{\Psi}\gamma_{\delta}\gamma^{5}\Psi = \frac{1}{2}\varepsilon^{\alpha\beta\gamma\delta}J_{\delta}^{0}$$
$$= 2|\rho|^{2}\varepsilon^{\alpha\beta\gamma\delta}A_{\delta}^{0} = 2\mathbf{A}^{\alpha} \bullet \mathbf{A}^{\beta} \times \mathbf{A}^{\gamma} \qquad (2.18)$$

The Yang-Mills version of the spin polarization tensor is easily shown from formula (2.11) to be:

$$S_{g}^{\alpha\beta\gamma} = \frac{1}{g}\text{Re}\,[A_{K}^{\alpha\beta}A^{K\gamma} - A_{K}^{\alpha\gamma}A^{K\beta}] \qquad (2.19)$$

In the limit of a large coupling constant g, the Yang-Mills formula (2.19) becomes using the definition of $A_{\alpha\beta}^{K} = \left(A_{\alpha\beta}^{0},\ \mathbf{A}_{\alpha\beta}\right)$ given in formula (2.12):

$$\lim_{g \to \infty} S_{g}^{\alpha\beta\gamma} = 2\mathbf{A}^{\alpha} \bullet \mathbf{A}^{\beta} \times \mathbf{A}^{\gamma} \qquad (2.20)$$

which equals $S^{\alpha\beta\gamma}$ by formula (2.18). Similarly, we can derive $T^{\alpha\beta}$ and $J^{\alpha}$ directly from the Yang-Mills formulas.

We mention in passing that, just as for Yang-Mills fields, the bispinor canonical (non-symmetric) energy-momentum tensor $T^{\alpha\beta}$ and spin polarization tensor $S^{\alpha\beta\gamma}$ satisfy a relation [23]:

$$\partial_\alpha S^{\alpha\beta\gamma} - T^{\beta\gamma} + T^{\gamma\beta} = 0 \qquad (2.21)$$

From this relation we can define a symmetric energy-momentum tensor, which is also conserved as follows:

$$\Theta^{\alpha\beta} = T^{\alpha\beta} + \frac{1}{2}\partial_\gamma(S^{\beta\gamma\alpha} + S^{\alpha\gamma\beta} - S^{\gamma\alpha\beta}) \qquad (2.22)$$

In general relativity, the symmetric tensor $\Theta^{\alpha\beta}$ is the bispinor source of the gravitational field, which is derived by varying the action with respect to the metric tensor [23]. (The action is formed of the Lagrangian (2.11) with the orthogonal constraint (2.10) expressed using Lagrange multiplyers.) Note that the general relativistic derivation of a symmetric energy-momentum tensor $\Theta^{\alpha\beta}$ is more self-evident using the Yang-Mills formulas rather than the bispinor formulas [24]. Also, for those interested in torsion theory generalizations, the interaction with torsion is much simpler to derive using the Yang-Mills formulas [7].

Although, as we have seen, embedding the gauge group SL(2,R) in the complex analytic group SL(2,C) simplifies derivations, for the Kaluza-Klein model presented in Section 3, it is more direct to express the Lagrangian (2.11) in terms of real gauge potentials $F_\alpha^K$ which are defined by setting:

$$j_\alpha^K = 4|\rho|^2 F_\alpha^K$$
$$s = 4|\rho|^2 \bar{\rho} \qquad (2.23)$$

Note from formulas (2.3) and (2.4) that the Noether currents $j_\alpha^K$ are real, and hence the gauge potentials $F_\alpha^K$ are also real. Also, note from formula (2.3) that the chiral U(1) gauge potential is $F_\alpha^3$. By formulas (2.5) and (2.23), these gauge potentials satisfy the orthogonal constraint:

$$F_\alpha^K F_{K\beta} = |\rho|^2 g_{\alpha\beta} \qquad (2.24)$$

In terms of the fields $(F_\alpha^K, \rho)$ the Lagrangian (2.11) becomes:

$$L_g = \frac{1}{4g} F_{\alpha\beta}^K F_K^{\alpha\beta} + \frac{1}{g_0} \overline{D_\alpha(\rho+\mu)} D^\alpha(\rho+\mu) \qquad (2.25)$$

where the Yang-Mills field tensor $F_{\alpha\beta}^L$ is given by:

$$F_{\alpha\beta}^L = \partial_\alpha F_\beta^L - \partial_\beta F_\alpha^L + g f_{JK}^L F_\alpha^J F_\beta^K \qquad (2.26)$$

and where we denote the SL(2,R) × U(1) Lie algebra structure constants as $f_{JK}^L$. Similar to formula (2.14), the covariant derivative $D_\alpha$ acts on the U(1) scalars $\rho$ and $\mu$ as follows:

$$D_\alpha(\rho+\mu) = \partial_\alpha \rho - i g_0 F_\alpha^3 (\rho+\mu) \qquad (2.27)$$

## 3. Kaluza-Klein radius smaller than the Planck length

In this section we will derive Dirac's bispinor Lagrangian (2.1) from a tetrad Kaluza-Klein model, which explicates both the orthogonal constraint (2.24) and the limit (2.15). The orthogonal constraint will be shown to be inherent in the structure of the tetrads, whereas the limit implies that the radius of the higher compact dimensions of the Kaluza-Klein model is vanishingly small compared with the Planck length, as a condition for the equality of the Einstein-Dirac and Kaluza-Klein Lagrangians.

To begin, we first describe the dynamical fields of the Kaluza-Klein tetrad model. Let M = X × G be the Kaluza-Klein manifold, with X a four dimensional space-time, and G the four-dimensional real Lie group SL(2,R) × U(1). On the space-time X, we assume the existence of a global, nonsingular tetrad of one-forms $\beta^K$ with K = 0, 1, 2, 3. The gravitational field on X, which we denote as $\beta$, is defined to be the unique metric tensor with the Minkowski signature, for which the tetrad $\beta^K$ is orthonormal, that is:

$$\beta = g_{JK} \beta^J \otimes \beta^K \qquad (3.1)$$

where

$$g_{JK} = g^{JK} = \begin{bmatrix} 1 & 0 & 0 & 0 \\ 0 & -1 & 0 & 0 \\ 0 & 0 & -1 & 0 \\ 0 & 0 & 0 & -1 \end{bmatrix} \qquad (3.2)$$

The tetrad of smooth one-forms $\beta^K$ uniquely determines its dual tetrad of smooth vector fields $b_K$ on X satisfying

$$\beta^K(b_J) = \delta_J^K \qquad (3.3)$$

where $\delta_J^K$ equals one if $J = K$, and equals zero otherwise. From formula (3.1), the vector fields $b_K$ form an orthonormal basis for each tangent space of X.

The fermion field on X we denote as ($F^K$, $\rho$), where $\rho$ is a complex scalar field and $F^K = |\rho|\beta^K$. Thus the dynamical fields are the tetrad of one forms $\beta^K$ and $\rho$. We will show that the gravitational field $\beta$ and the bispinor field $\Psi$ (which together have 10 + 8 = 18 real components), are represented faithfully by $\beta^K$ and $\rho$ (which also have 16 + 2 = 18 real components) [2]. We will then derive the usual Einstein-Dirac Lagrangian from the Kaluza-Klein Lagrangian (3.22) for the fields $\beta^K$ and $\rho$.

On G, the four-dimensional real Lie group SL(2,R) × U(1), we fix a nonsingular tetrad of right-invariant one-forms $\alpha^K$ with $K = 0, 1, 2, 3$. The tetrad of right-invariant one-forms $\alpha^K$ defines a right-invariant metric on the Lie group G given by:

$$\alpha = g_{JK}\, \alpha^J \otimes \alpha^K \qquad (3.4)$$

where $g_{JK}$ has the same form as the Minkowski metric in the definition (3.2). Since G is a four-dimensional Lie group, the $\alpha^K$ form a basis for the dual of the Lie algebra of G.

For vector fields v and w on G, we will denote the inner product with respect to the metric $\alpha$ by $\langle v, w \rangle$, that is:

$$\langle v, w \rangle = \alpha(v, w) = g_{JK}\, \alpha^J(v)\, \alpha^K(w) \qquad (3.5)$$

The tetrad of right-invariant one-forms $\alpha^K$ uniquely determines a dual tetrad of right-invariant vector fields $a_K$ on G satisfying

$$\alpha^K(a_J) = \delta_J^K \qquad (3.6)$$

The right-invariant vector fields $a_K$ form a basis for the Lie algebra of G. This basis is orthonormal, since from formulas (3.5) and (3.6) we get:

$$\langle a_J, a_K \rangle = g_{JK} \quad (3.7)$$

We can choose the fixed tetrad $\alpha^K$ so that the vector fields $a_K$ satisfy the following SL(2,R) × U(1) commutation relations:

$$[a_0, a_1] = -\delta^{-1} a_2$$

$$[a_0, a_2] = \delta^{-1} a_1 \quad (3.8)$$

$$[a_1, a_2] = \delta^{-1} a_0$$

where $\delta$ is a length parameter. All other commutators vanish. As usual in general relativity, both length and time carry the same unit. As on any physical manifold, the one-forms $\alpha^K$ carry units of length, so that their duals, the vector fields $a_K$ in formula (3.8), carry units of mass (i.e., inverse length). From formulas (3.7) and (3.8) it is evident that $\delta$ is the radius of the U(1) subgroups of SL(2,R). Formula (3.8) can be written more succinctly as:

$$[a_J, a_K] = \frac{1}{\delta} f_{JK}^L a_L \quad (3.9)$$

which defines the Lie algebra structure constants $f_{JK}^L$. Note that the structure constants $f_{JK}^L$ are dimensionless, so that the length parameter $\delta$ is required in formula (3.9) to balance the dimensions. Also, in formula (3.4), the metric constants $g_{JK}$ are dimensionless. Although we do not make use of the following property in the tetrad Kaluza-Klein model, note from formulas (3.2) and (3.8) that $f_{JKL} = g_{LM} f_{JK}^M$ is completely antisymmetric in the indices J, K, and L. When this property holds, the metric is called "bi-invariant", since it is both right and left invariant [25]. We will see generally that the tetrad Kaluza-Klein model does not require that the right-invariant metric $\alpha$ given in formula (3.4) be bi-invariant.

Note that while the orthonormal and commutation relations (3.7) and (3.8) determine the radius of the U(1) subgroups of SL(2,R), they do not determine the radius of the U(1) factor of the Lie group G = SL(2,R) × U(1). The radius of the U(1) factor of G will be denoted as $\delta_0$. The ratio $\delta/\delta_0$ is a parameter which we can equate to the ratio $g_0/g$ of coupling constants in the Yang-Mills Lagrangian (2.25). That is, the length parameters $\delta_0$ and $\delta$ of the tetrad Kaluza-Klein model will be set as $\delta_0 = (2/3)\delta$ in correspondence with $g_0 = (3/2)g$ in the Lagrangian (2.25).

Thus on the Kaluza-Klein manifold $M = X \times G$, we can define a fixed tetrad of one-forms $\alpha^K$ and a dynamic tetrad of one-forms $\beta^K$ induced from the projections of M onto its factors G and X. ($\alpha^K$ and $\beta^K$ on M are the pullbacks of $\alpha^K$ on G and $\beta^K$ on X by the projection maps.) We define a third tetrad of one-forms $\nu^K$ on M by:

$$\nu^K = \alpha^K - (\kappa\delta)^{1/3}|\rho|\beta^K \qquad (3.10)$$

where $\kappa$ is $16\pi/3$ times Newton's constant $\kappa_0$, and $\rho$ is a complex scalar field on M. Note that the constant $\kappa$ has dimension of length squared, the constant $\delta$ has dimension of length as in formula (3.9), the scalar field $\rho$ has dimension of mass, and the one-forms $\alpha^K$, $\beta^K$, and $\nu^K$ each have dimension of length.

The one-forms $(\beta^K, \nu^K)$ form a basis for each cotangent space of $M = X \times G$. The Kaluza-Klein metric on M is defined to be:

$$\gamma = g_{JK}(\beta^J \otimes \beta^K + \nu^J \otimes \nu^K) \qquad (3.11)$$

which depends only on the dynamical fields $\beta^K$ and $\rho$, since $\alpha^K$ in formula (3.10) is fixed by the basis chosen for the Lie algebra of G.

To demonstrate that $\gamma$ is a Kaluza-Klein metric, we define local coordinate one-forms $dx^\alpha$ with $\alpha = 0,1,2,3$ on an open chart $V \subset X$. The gravitational field $\beta$ is expressed locally on V by:

$$\beta = g_{\alpha\beta}\, dx^\alpha \otimes dx^\beta \qquad (3.12)$$

Writing $\beta^K = \beta^K_\alpha\, dx^\alpha$, we obtain from formulas (3.1) and (3.12):

$$g_{\alpha\beta} = g_{JK}\, \beta^J_\alpha\, \beta^K_\beta \qquad (3.13)$$

If we choose $(dx^\alpha, \alpha^K)$ for a basis of one-forms, then from formulas (3.10) and (3.13), the Kaluza-Klein metric (3.11) has the following components:

$$\gamma = \begin{bmatrix} g_{\alpha\beta} + \lambda^2 g_{JK} F_\alpha^J F_\beta^K & -\lambda F_\alpha^J g_{JK} \\ -\lambda g_{JK} F_\beta^K & g_{JK} \end{bmatrix} \quad (3.14)$$

where $\lambda = (\kappa\delta)^{1/3}$ is a Kaluza-Klein parameter having dimension of length [14], and

$$F_\alpha^K = |\rho| \beta_\alpha^K \quad (3.15)$$

Thus, $\gamma$ is precisely the Kaluza-Klein metric [14] for the gravitational field $g_{\alpha\beta}$ and the gauge potentials $F_\alpha^K$. By formulas (3.13) and (3.15) the $F_\alpha^K$ satisfy:

$$g_{JK} F_\alpha^J F_\beta^K = |\rho|^2 g_{\alpha\beta} \quad (3.16)$$

which is precisely the orthogonal constraint (2.24). Furthermore, by formula (3.13), the gravitational field $g_{\alpha\beta}$ has the same (Minkowski) signature as $g_{JK}$ on G.

We denote the vector fields dual to $(\beta^K, \alpha^K)$ as $(b_K, a_K)$. The vector fields dual to $(\beta^K, v^K)$ are then $(v_K, a_K)$, where from formula (3.10):

$$v_K = b_K + (\kappa\delta)^{1/3} |\rho| a_K \quad (3.17)$$

From formula (3.11), the vector fields $(v_K, a_K)$ form an orthonormal basis with respect to the Kaluza-Klein metric $\gamma$ on each tangent space of M.

We extend the inner product notation in formula (3.5) to vector fields v and w defined on M as follows:

$$\langle v, w \rangle = \gamma(v, w) = g_{JK} [\beta^J(v)\beta^K(w) + v^J(v)v^K(w)] \quad (3.18)$$

Thus, for the orthonormal vector fields $v_K$ and $a_K$ defined on M:

$$\langle v_J, v_K \rangle = \langle a_J, a_K \rangle = g_{JK}$$
$$\langle v_J, a_K \rangle = 0 \quad (3.19)$$

for all indices J, K = 0, 1, 2, 3. That is, with respect to the basis $(v_K, a_K)$, the Kaluza-Klein metric $\gamma$ becomes:

$$\gamma = \begin{bmatrix} g_{JK} & 0 \\ 0 & g_{JK} \end{bmatrix} \qquad (3.20)$$

The manifold $M = X \times G$ has a natural right action of G defined by $h(x, g) = (x, gh)$ for each $(x, g) \in M$ and $h \in G$. For $v_K$ to be right invariant, it is necessary and sufficient that $b_K$ and $|\rho|$ depend only on the space-time coordinates $x \in X$. Specifically, we assume that the complex scalar field $\rho$ has the form:

$$\rho = e^{iy/\delta_0} \tilde{\rho}(x) \qquad (3.21)$$

where y is a global U(1) coordinate of G for which $a_3 = -\partial/\partial y$ is a U(1) unit vector field on G which commutes with every right-invariant vector field on G (see formulas (3.7) and (3.8)).

Our goal in this section is to derive the Einstein and Dirac Lagrangians from the following Lagrangian for the fields $(\beta^K, \rho)$:

$$L = \frac{1}{16\pi\kappa_0} R_v + \frac{1}{g_0} \overline{v_K(\rho+\mu)} \, v^K(\rho+\mu) \qquad (3.22)$$

where $\kappa_0$ and $g_0$ are constants ($\kappa_0$ is Newton's constant), and where $v^K = g^{JK} v_J$. The mass parameter $\mu$ is defined on M by:

$$\mu = e^{iy/\delta_0} \tilde{\mu} \qquad (3.23)$$

where $\tilde{\mu}$ is a constant. $R_v$ is the sum of sectional curvatures over the four-dimensional subspaces spanned by the orthonormal tetrad $v_K$ in each tangent space of M:

$$R_v = g^{JK} g^{LM} \langle R(v_J, v_L) v_K, v_M \rangle \qquad (3.24)$$

where R( , ) is the curvature two-form [25] associated with the Kaluza-Klein metric $\gamma$ on M.

Let $d\gamma$ denote the volume form on $M = X \times G$ defined by the Kaluza-Klein metric $\gamma$. (We do not confuse the symbol "d" with exterior differentiation since the metric $\gamma$ is not a differential form.) Similarly let $d\alpha$ and $d\beta$ denote the volume forms defined by the metrics $\alpha$ and $\beta$ on the manifolds G and X, respectively. Note that $d\alpha$ is a fixed volume form on G, whereas $d\beta$ depends on the dynamic fields $\beta^K$. Since the one-forms $(\beta^K, v^K)$ are orthonormal, we see from formula (3.10) that

$$d\gamma = d\beta \wedge d\alpha \qquad (3.25)$$

Therefore, the action associated with the Lagrangian (3.22) is given by:

$$S = \int L(\beta^K, \rho) \, d\beta \wedge d\alpha \qquad (3.26)$$

Note that in the action (3.26), the gravitational field $g_{\alpha\beta}$ and the bispinor field $\Psi$, which together have $10 + 8 = 18$ real components, are represented by $\beta^K$ and $\rho$, which also have $16 + 2 = 18$ real components [2].

We show in the following theorem that the Lagrangian (3.22) equals the Hilbert-Einstein Lagrangian for the gravitational field plus the Dirac-Yang-Mills Lagrangian (2.25). The constraint (2.24) of the Dirac-Yang-Mills equation has already been shown to be a consequence of the tetrad in formula (3.16).

**THEOREM**: If we define the constants $\kappa$, $g$, $g_0$ in terms of Newton's constant $\kappa_0$ and the length parameters $\delta$ and $\delta_0$ as follows:

$$\kappa = \frac{16\pi}{3}\kappa_0$$

$$g = \frac{(\kappa\delta)^{1/3}}{\delta} \qquad (3.27)$$

$$g_0 = \frac{(\kappa\delta)^{1/3}}{\delta_0}$$

then the total Lagrangian L given in formula (3.22) equals the Hilbert-Einstein Lagrangian $(16\pi\kappa_0)^{-1} R_X$ for the gravitational field plus the Dirac-Yang-Mills Lagrangian $L_g$ given in formula (2.25) for the fermion field, and similarly for the total action (3.26). Furthermore, the limit (2.15) required to obtain Dirac's bispinor equation, forces the length parameters $\delta$ and $\delta_0$ in the Kaluza-Klein model to become vanishingly small compared with the Planck length $\lambda_P = \kappa_0^{1/2}$.

**PROOF**: We will derive an alternative local expression for the Lagrangian (3.22), which simplifies the computations. Define a local coordinate tetrad $v_\alpha$ as follows:

$$v_\alpha = \partial_\alpha + (\kappa\delta)^{1/3} F_\alpha^K a_K \qquad (3.28)$$

Since $v_\alpha = \beta_\alpha^K v_K$, the tetrads $v_K$ and $v_\alpha$ in formulas (3.17) and (3.28) span the same four dimensional distribution over the Kaluza-Klein manifold M.

The inverse relation, $v_K = b_K^\alpha v_\alpha$, where $b_K^\alpha$ are the components of the vector fields $b_K = b_K^\alpha \partial_\alpha$, follows from formulas (3.3), (3.15), (3.17), and (3.28). Similarly, formulas (3.3) and (3.13) imply:

$$b_K^\beta = g_{JK} g^{\alpha\beta} \beta_\alpha^J$$
$$g^{\alpha\beta} = g^{JK} b_J^\alpha b_K^\beta \qquad (3.29)$$

Then, substituting $v_K = b_K^\alpha v_\alpha$ into $R_v$, the sum of sectional curvatures over the distribution spanned by $v_K$ in formula (3.24), gives:

$$R_v = g^{\alpha\beta} g^{\gamma\delta} \langle R(v_\alpha, v_\gamma) v_\beta, v_\delta \rangle \qquad (3.30)$$

and the Lagrangian (3.22) equals:

$$L = \frac{1}{16\pi\kappa_0} R_v + \frac{1}{g_0} \overline{v_\alpha(\rho+\mu)}\, v^\alpha(\rho+\mu) \qquad (3.31)$$

Formula (3.30) is evaluated by computing $R_v$ using the vector fields $(v_\alpha, a_K)$ as a basis on M. Note that with respect to this basis, the Kaluza-Klein metric (3.11) has the following components:

$$\gamma = \begin{bmatrix} g_{\alpha\beta} & 0 \\ 0 & g_{JK} \end{bmatrix} \qquad (3.32)$$

The local expressions of $v_\alpha$, $R_v$, and $\gamma$ given in formulas (3.28), (3.30), and (3.32) are equal to the usual expressions in Kaluza-Klein theory [14]. A straightforward derivation using the commutation relations (3.9) shows that:

$$R_v = R_X + \frac{3}{4}(\kappa\delta)^{2/3} F_{\alpha\beta}^K F_K^{\alpha\beta} \qquad (3.33)$$

where $R_X$ denotes the scalar curvature of X, and

$$F_{\alpha\beta}^{K} = \partial_{\alpha} F_{\beta}^{K} - \partial_{\beta} F_{\alpha}^{K} + g f_{MN}^{K} F_{\alpha}^{M} F_{\beta}^{N} \qquad (3.34)$$

where $\partial_{\alpha}$ are the coordinate vector fields dual to $dx^{\alpha}$ in formula (3.12) and $g = (\kappa/\delta^2)^{1/3}$. Note in formula (3.33) that indices are raised and lowered in the obvious way. That is

$$F_{K}^{\alpha\beta} = g^{\gamma\alpha} g^{\delta\beta} g_{JK} F_{\gamma\delta}^{J} \qquad (3.35)$$

(Because in formula (3.24), we restricted $R_v$ to the tetrad $v_K$, the scalar curvature of G does not occur in formula (3.33)).

Having computed $R_v$ in formula (3.33), and choosing the constants $\kappa$, $g$, and $g_0$ as in formula (3.27), we see in formula (3.31) that the total Lagrangian L equals the Hilbert-Einstein Lagrangian for $g_{\alpha\beta}$ plus the Dirac-Yang-Mills Lagrangian $L_g$ given in formula (2.25) for $F_{\alpha}^{K}$ and $\rho$.

Furthermore, since $\delta = (3/2)\delta_0$, formula (3.27) gives:

$$\delta_0 = \left(\frac{8\pi}{g_0^3}\right)^{1/2} \lambda_P \qquad (3.36)$$

which relates the Kaluza-Klein radius $\delta_0$ to the Planck length $\lambda_P = \kappa_0^{1/2}$. Thus in the limit required to obtain Dirac's equation, that is, as $g_0$ becomes infinitely large, $\delta_0$ must become vanishingly small compared to the Planck length. The same is true for the radius $\delta = (3/2)\delta_0$. **Q. E. D.**

The following observations derive from the proof of the theorem and demonstrate that the Lagrangian (3.22) significantly generalizes the Kaluza-Klein theory. First, from formulas (3.10), (3.11), (3.13), and (3.15), and from the local expressions of the Lagrangian in formulas (3.28), (3.30), and (3.31), the gauge group G of the Kaluza-Klein manifold $M = X \times G$ can be generalized to larger Lie groups of dimension $d > 4$. For such generalizations we define $v_K = b_K^{\beta} v_{\beta}$ where $b_K^{\beta} = g^{\alpha\beta} g_{JK} \beta_{\alpha}^{J}$. Although, the d global vector fields $v_K$ are too many to form a tetrad when $d > 4$, they span a four dimensional distribution (spanned locally by the coordinate tetrads $v_{\alpha}$).

Second, the nonphysical cosmological constant, which is the scalar curvature (denoted as $R_G$) of the Lie group G occurring in the Lagrangian of the usual Kaluza-Klein model [14], is absent in the Lagrangian (3.22) because in formula (3.24) we restricted $R_v$ to the four dimensional distribution spanned by the $v_K$.

Third, for the same reason, even though the metric given in formula (3.4) is bi-invariant (i.e., both right and left invariant), the theorem does not require that the right-invariant metric to also be left-invariant, which in the usual Kaluza-Klein model restricts the choice of Lie groups[14].

## 4. Measurement of the Kaluza-Klein radius

In this section we will first show that exact plane wave solutions (in a Minkowski space-time) of the Euler-Lagrange equations for the Lagrangian (2.11) with the orthogonal constraint (2.10), are in one-to-one correspondence with the plane wave solutions of Dirac's bispinor equation (which as previously shown is obtained in the limit that the Yang-Mills coupling constant $g_0$ (and $g = (2/3) g_0$) becomes infinite). Using a wave packet approximation, we derive quasi-linear partial differential equations for wave packets with slowly varying amplitude and momentum. Solutions to these equations propagate along two families of characteristic curves with a small velocity difference. Thus a single fermion wave packet may split into two wave packets traveling with slightly different velocities.

Since the velocity splitting depends on the coupling constant $g_0$ (and vanishes as $g_0$ becomes infinite), measurement of the velocity splitting will determine $g_0$ and through formula (3.36), the Kaluza-Klein radius $\delta_0$. We discuss experiments to measure such velocity splitting and show that currently achievable experiments could bound $\delta_0$ to less than $10^{-25}$ times the Planck length if velocity splitting were not observed.

The Euler-Lagrangian equations for (2.10) and (2.11) have exact plane wave solutions of the form [16]:

$$A_\alpha^0(x^\beta) = A_\alpha^0(0)$$

$$\mathbf{A}_\alpha(x^\beta) = e^{2i\theta(x^\beta)T} \mathbf{A}_\alpha(0) \quad (4.1)$$

$$\rho(x^\beta) = \rho(0)$$

where $x^\beta \in R^4$ denotes the space-time coordinates, T generates a one parameter subgroup of SL(2,C) gauge transformations, and $\theta(x^\beta) = p_\beta x^\beta$ where $p_\beta \in R^4$ denotes the momentum variables. Note that if $A_\alpha^K(0)$ and $\rho(0)$ satisfy the orthogonal constraint (2.10), then the same is true for $A_\alpha^K(x^\beta)$ and $\rho(x^\beta)$ for all $x^\beta \in R^4$, since in formula (4.1), the SL(2,C) gauge transformations generated by T preserve the orthogonal constraint. Note also that

$$T(\mathbf{A}_\alpha) = i\boldsymbol{\omega} \times \mathbf{A}_\alpha \qquad (4.2)$$

for some $\boldsymbol{\omega} \in C^3$ satisfying $\boldsymbol{\omega} \cdot \boldsymbol{\omega} = 1$. (The reader is reminded that SL(2,C) is the complexification of SU(2) for which we can take $\boldsymbol{\omega} \in R^3$.)

Differentiating formula (4.1), we get using (4.2):

$$\partial_\alpha A_\beta^0 = 0$$

$$\partial_\alpha \mathbf{A}_\beta = -2p_\alpha \boldsymbol{\omega} \times \mathbf{A}_\beta \qquad (4.3)$$

$$\partial_\alpha \rho = 0$$

Note in formula (4.3) that the $\mathbf{A}_\alpha$ have twice the rotation rate of bispinors, and $p^\alpha p_\alpha = m^2$ where m is the mass of the plane wave solution (4.1). We also assume that the plane waves (4.1) satisfy the same conditions which are satisfied by bispinor plane waves, given as follows:

$$p^\alpha A_\alpha^0 = 0$$
$$\qquad (4.4)$$
$$p^\alpha \mathbf{A}_\alpha = \pm m |\rho| \boldsymbol{\omega}$$

where the positive sign is used for particles and the negative sign for antiparticles. Since $\rho$ is constant by (4.3), formula (4.4) can be regarded as the initial conditions for the fields $A_\alpha^K$. Note that formula (4.4) is consistent with $p^\alpha p_\alpha = m^2$, as well as $\boldsymbol{\omega} \cdot \boldsymbol{\omega} = 1$ and the orthogonal constraint (2.10). Moreover, $p_\alpha$ for particles becomes $-p_\alpha$ for antiparticles. Conversely, with $p_\alpha$ so defined, formula (4.4) defines $\boldsymbol{\omega}$, and hence the gauge generator T in formulas (4.1) and (4.2).

Since in the following we only deal with particles (e.g. electrons and protons), we will disregard the antiparticle formulas. Since $\rho$ is constant we can choose $\rho > 0$ for particle plane waves. (This is equivalent to choosing a positive mass parameter $\mu$ in the Lagrangian (2.11).)

Note from the orthogonal constraint (2.10) that if we choose $\mu > 0$ and $\rho > 0$ for particle plane waves, then each plane wave (4.1) has a constant amplitude equal to $\rho$. As shown in previous work [16], each exact plane wave solution (4.1) has a mass $m = m(\rho)$, which depends on its constant amplitude $\rho$. To derive $m(\rho)$, substitute the plane wave (4.1) into the Euler-Lagrange equations for the Lagrangian (2.11) with the orthogonal constraint (2.10) expressed using Lagrange multiplyers [16]. Then, the mass m can be shown to satisfy the following quadratic equation:

$$(m - m_0)^2 + b(m - m_0) + c = 0 \tag{4.5}$$

where

$$m_0 = \frac{1}{2} g_0 \mu$$

$$b = 2 m_0 + g_0 \rho \tag{4.6}$$

$$c = \frac{1}{3} m_0^2$$

We deduce from formulas (4.5) and (4.6) that the mass m is positive for all positive amplitudes $\rho$. Expanding m in powers of $m_0 / g_0 \rho$, we have approximately for a large Yang-Mills coupling constant $g_0$ that

$$m \approx m_0 - \frac{m_0^2}{3 g_0 \rho} \tag{4.7}$$

Note that in the limit as $g_0$ becomes infinitely large, the mass m becomes $m_0$, which is constant, and hence independent of the amplitude $\rho$.

Wave packets are defined to be plane waves with slowly varying parameters (e.g. amplitude, spin, and momentum). To describe such wave packets we introduce "slow" coordinates $y^\beta = \varepsilon x^\beta$, where $\varepsilon > 0$ is a small parameter, into formula (4.1) by the substitutions $A_\alpha^K(y^\beta)$ for $A_\alpha^K(0)$, $\rho(y^\beta)$ for $\rho(0)$, $\varepsilon^{-1} \theta(y^\beta)$ for $p_\alpha x^\alpha$, and:

$$\partial_\alpha = \frac{\partial}{\partial x^\alpha} = \varepsilon \frac{\partial}{\partial y^\alpha}$$

$$p_\alpha = \frac{\partial \theta}{\partial y^\alpha} \tag{4.8}$$

Using the Whitham method [15], we express the Lagrangian (2.11) in terms of $\varepsilon$, $y^\beta$, $\theta$, $A_\alpha^K$, and $\rho$. Then, because the Lagrangian (2.11) and the orthogonal constraint (2.10) are independent of the phase $\varepsilon^{-1}\theta$ in formula (4.1), we may set $\varepsilon = 0$ and drop the distinction between $x^\beta$ and $y^\beta$. The resulting Euler-Lagrange equations are the equations governing the wave packets [15], and are given by:

$$p^\alpha p_\alpha = m^2$$

$$\partial_\alpha p_\beta = \partial_\beta p_\alpha \qquad (4.9)$$

$$\partial_\alpha J^\alpha = 0$$

where $m = m(\rho)$ is given in formula (4.5), and

$$J_\alpha = J u_\alpha = \left(\frac{12m}{g_0}\rho^2 + 4\rho^3\right) u_\alpha \qquad (4.10)$$

where $u_\alpha = p_\alpha/m$. $J_\alpha$ is the electric current (i.e., the Noether current which is gauge parallel to $\boldsymbol{\omega}$ in formula (4.2)). The first two equations (4.9) are called the eikonal equations, and the last equation (4.9) expresses the conservation of the electric current $J_\alpha$.

To analyze equation (4.9), we now consider a space-time with one space dimension such that,

$$u^\alpha = (u^0, u^1) = \frac{(1, u)}{\sqrt{1-u^2}} \qquad (4.11)$$

for velocity parameter u, and similarly, we denote $x^\alpha = (x, t)$. Formula (4.9) becomes:

$$\frac{\partial}{\partial t}(Ju^0) + \frac{\partial}{\partial x}(Ju^1) = 0$$

$$\frac{\partial}{\partial t}(mu^1) + \frac{\partial}{\partial x}(mu^0) = 0 \qquad (4.12)$$

where the two dependent variables are the amplitude $\rho$ and the velocity u. (Recall from formulas (4.7), (4.10), and (4.11) that $m = m(\rho)$, $J = J(\rho)$, and $u^\alpha = u^\alpha(u)$ for $\alpha = 0, 1$.)

Characteristic curves for the quasi-linear partial differential equations (4.12) are easily derived [15], and are given by:

$$\frac{dx}{dt} = \frac{u \pm \Delta}{1 \pm u\Delta} \tag{4.13}$$

where

$$\Delta = \sqrt{\frac{Jm'}{J'm}} \tag{4.14}$$

where $J'$ and $m'$ denote the derivatives of J and m with respect to $\rho$. On the two families of characteristic curves, signified by $\pm$ signs in formula (4.13), we have

$$\frac{1}{1-u^2} du = \pm \frac{J'\Delta}{J} d\rho \tag{4.15}$$

Note that for the two families of characteristic curves (4.13), the group velocity $dx/dt$ equals the relativistic addition of velocities, u and $\pm \Delta$, respectively. The velocity splitting is defined to be $2\Delta$. Substituting the expressions for m and J from formulas (4.7) and (4.10) into formula (4.14), we derive the following approximate formula for velocity splitting for a large Yang-Mills coupling constant $g_0$:

$$2\Delta \approx \sqrt{\frac{m_0}{g_0 \rho}} \tag{4.16}$$

Here and henceforth we will ignore non-significant factors (e.g., 2/3 in formula (4.16)) in deriving approximate formulas.

Consider a fermion wave packet of length $L_0$, volume $L_0^3$, and density $1/L_0^3$. For a wave packet of nearly uniform density this implies $\rho \approx 1/L_0$ (see formulas (2.7) and (2.10)). To minimally detect velocity splitting over the free path of the fermion, the original wave packet must split into two wave packets separated by a distance of at least $L_0$. Thus, we must have $L_0 \approx 2t_1\Delta$ where $t_1 = L_1/u$ is the time for the fermion to travel the length of the free path $L_1$ at the fermion velocity u. Hence $2\Delta \approx uL_0/L_1$. Substituting $2\Delta \approx uL_0/L_1$ and $\rho \approx 1/L_0$ into formula (4.16), we get approximately:

$$g_0 \approx \frac{m_0 L_1^2}{u^2 L_0} \tag{4.17}$$

Then, substituting $g_0$ in formula (4.17) into (3.36) gives the following approximate relation between the Kaluza-Klein radius $\delta_0$ and experimentally determined parameters (the fermion mass $m_0$, fermion velocity u, wave packet length $L_0$, and free path length $L_1$) for which velocity splitting is minimally detectable:

$$\frac{\delta_0}{\lambda_P} \approx \left(\frac{u^2 L_0}{m_0 L_1^2}\right)^{3/2} \tag{4.18}$$

Consider a current experiment [17] where single electrons are emmited at 100 km intervals in wave packets of length $10^{-5}$ m, traveling a free path of 1m at half the speed of light. That is, $L_0 = 10^{-5}$ m, $L_1 = 1$ m and $u = .5$. Assuming that velocity splitting is not observed, and given that the electron mass is $4 \times 10^{11} \text{m}^{-1}$, we deduce from formulas (4.17) and (4.18) that $g_0$ must be greater than $10^{17}$, and hence $\delta_0$ must be smaller than $10^{-25}$ times the Planck length $\lambda_P$. Since in formula (4.18) the estimate of $\delta_0$ is proportional to the velocity u cubed and inversely proportional to the 3/2 power of the mass $m_0$, refined experiments using slower electrons or (more massive) protons could improve the bound on $\delta_0$ by twenty orders of magnitude.